\documentclass[12pt]{article}
\usepackage{graphicx}

\begin{document}
\begin{titlepage}
%
%






{\ }

\begin{center}
\centerline{Solid state physics of impact crater formation: a detail}

\end{center}
\begin{center}
\centerline{V.Celebonovic}			
\end{center}
\vskip5mm

\begin{center}
\centerline{Institute of Physics,University of Belgrade,Pregrevica 118,11080 Zemun-Belgrade,Serbia}
\vskip3mm 
\centerline{vladan@ipb.ac.rs}
\end{center}



\begin{abstract}
Impact craters exist on solid surface planets, their satellites and many asteroids.The aim of this paper is to propose a theoretical expression for the product $\rho r^{3} v_{1}^{2}$,where the three symbols denote the mass density,radius and speed of the impactor. The expression is derived using well known results of solid state physics,and it can be used in estimating parameters of impactors which have led to formation of craters on various solid bodies in the Solar System.
\end{abstract}
\end{titlepage}
\section{Introduction}
Craters of various sizes have been observed on the terrestrial planets,their satellites and the major part of the asteroids.The study of these craters,and the resulting constraints on the related impactors has become a separate field of research in planetary science,see for example
\cite{[1]}. One of the fundamental questions concerning the impactors is what can be concluded about them by combining astronomical data with results of solid state physics. 

Recent theoretical work ( for example \cite{[2]}) has shown that the application of elementary principles of solid state physics to this problem gives physically plausible results in reasonable agreement with those obtained by celestial mechanics. It was assumed in \cite{[2]} that the material of the target was a crystal lattice,and the calculations were performed {\it per unit volume}. It was assumed there the condition for the formation of a crater is that the kinetic energy of a unit volume of the impactor has to be equal to or greater than the internal energy of a unit volume of the material of the target. The result was an expression for the minimal speed which an impactor of given parameters must have when hitting a target with a predefined set of parameters, in order to form a crater. 

The aim of the calculation reported here is to take into account the dimensional effects - to consider both the impactor and the crater it forms as objects of finite dimensions. The novelty of the approach discussed in the next section, compared to existing work such as \cite{[3]},is the generality - it is based on principles of solid state physics,and it can be applied to any solid material. Formation of impact craters is here discussed from the viewpoint of pure solid state physics as the following analogous problem: what kinetic energy of an impactor is needed to produce a hole of given dimensions in a material with a predefined set of parameters?
\section{Calculations}  
 The physical keypoint of this calculation is the idea that the kinetic energy of the impactor must be equal to or greater than the internal energy of some volume,$V_{2}$, of the target. The kinetic energy of the impactor of mass $m_{1}$ and speed $v_{1}$ is 
\begin{equation}
	E_{k}=\frac{1}{2}m_{1} v_{1}^{2} = \frac{1}{2} m_{1} (vcos\theta)^{2}
\end{equation}
where $\theta$ is the angle of the trajectory of the impactor immediately preceeding the impact with respect to the vertical at the point of impact.The internal energy of a volume $V_{2}$ of a solid is \cite{[4]}
\begin{equation}
	E_{2}=\frac{\pi^{2}}{10}\frac{(k_{B} T_{2})^{4}}{(\hbar \bar{V})^{3}}V_{2}
\end{equation}
all the symbols have their standard meanings and $\bar{V}$ is the speed of ellastic waves in the material of the target.One of the problems consists in knowing the volume $V_{2}$,which is,in first approximation,equal to the volume of the crater. It is known from theoretical work \cite{[3]} and references given there that the form of a crater is determined by a combination of "gravity scaling" and "strength scaling", where the term "strength" reffers to the material strength of the target. Explicite expressions for the form of craters exist in \cite{[3]} and they are applicable to 4 material types. In order to simplify somewhat the calculations,and get slightly more general results, it was assumed in the present work that craters have the shape of a half of a rotating elipsoid,with distinct semi axes,denoted by $a$,$b$ and $c$. Physically,$a$ and $b$ denote the semi axes of the "opening" of the crater,while $c$ is the depth. In that approximation,the volume of the crater is  obviously given by
\begin{equation}
V_{2}= \frac{2}{3}\pi a b c
\end{equation}
Inserting eq.(3) into eq.(2) leads to 
\begin{equation}
	E_{2}= \frac{\pi^{3}}{15}\frac{(k_{B} T_{2})^{4}}{(\hbar \bar{V})^{3}} a b c
\end{equation}
Inserting the definition of the mass density $(\rho_{1}= \frac{m_{1}}{V_{1}})$ in eq.(1) and assuming further that the impactor has the form of a sphere of radius $r_{1}$,the following expression for the kinetic energy of the impactor is obtained:
\begin{equation}
	E_{1}=\frac{1}{2}m_{1}v_{1}^{2} = \frac{1}{2}\times\frac{4}{3}\pi r_{1}^{3}\rho_{1}(v\cos\theta)^{2}=\frac{2}{3}\pi\rho_{1}r_{1}^{3}(v\cos\theta)^{2}
\end{equation}
Equating expressions (5) and (4) and grouping terms, gives the following expression for the velocity which  the impactor with mass density $\rho_{1}$ and radius $r_{1}$ must have in order to form a crater of depth $c$ and semi axes of the "opening" $a$ and $b$ in a target with temperature $T_{2}$
\begin{equation}
	\rho_{1}r_{1}^{3}(v\cos\theta)^{2}=\frac{\pi^{2}}{10}\frac{(k_{B} T_{2})^{4}}{(\hbar \bar{V})^{3}} a b c 
\end{equation}
This expression can be reformulated as
\begin{equation}
	\rho_{1}r_{1}^{3}(v\cos \theta)^{2}= \frac{\pi^{2}}{10} \frac{(k_{B} T_{2})^{4}}{(\hbar)^{3}} (\frac{\partial P}{\partial \rho})^{-3/2} a b c 
\end{equation}

Equation (6) has the advantage of grouping known or measurable physical quantities on the right hand side,and those which are unknown on the left side. 
The application of eq.(7) demands the knowledge of the equation of state (EOS) of the material of the target. Generally speaking,regardless of its detailed form, an EOS can be expressed in the following analytical form
\begin{equation}
	P(\rho)=\sum_{i=0}^{\infty}a_{i} (\frac{\rho}{\rho_{0}})^{i}
\end{equation}
where $a_{i}$ are some coefficients, $\rho_{0}$ is the density at some pressure $P_{0}$ and $\rho$ is the density at pressure $P$. Limiting this developement to first order terms, it follows that
\begin{equation}
	(\frac{\partial P}{\partial \rho})^{- 3/2}\cong(\frac{a_{1}}{\rho_{0}})^{-3/2}[1-3\frac{a_{2}}{a_{1}}\frac{\rho}{\rho_{0}}]
\end{equation}
A well known example of the EOS of a solid is the Birch-Murnaghan EOS \cite{[5]}
\begin{eqnarray}
	P(\rho)=\frac{3B_{0}}{2}\left[(\frac{\rho}{\rho_{0}})^{7/3}-\frac{\rho}{\rho_{0}})^{5/3}\right]\times\nonumber\\
	\left\{1+(3/4)(B_{0}'-4)\left[(\rho/\rho_{0})^{2/3}-1\right]\right\}
\end{eqnarray}
where $B_{0}=-V(\frac{\partial P}{\partial V})_{T}$ is the bulk modulus of the material and $B_{0}'=(\frac{\partial B}{\partial P})_{T}$ is its pressure derivative. The symbols $\rho_{0}$ and $\rho$ denote the mass density of a specimen under consideration at the initial value of the pressure $P_{0}$ and at some arbitrary value $P$. 
The final form of eq.(7) assuming that the material of the target can be described by the Birch Murnaghan EOS is
\begin{eqnarray}
\rho_{1}r_{1}^{3}(vcos \theta)^{2}= \nonumber\\
\frac{8 \pi^{2} \sqrt{2}}{5} \frac{(k_{B} T_{2})^{4}}{\hbar^{3}}\frac{\rho_{0}^{2}}{B_{0}^{3/2}}abc[27(B_{0}'-4)\rho^{2}+\nonumber\\14(14-3B_{0}')\rho_{0}\rho(\rho/\rho_{0})^{1/3}\nonumber\\
+5(3B_{0}'-16)\rho_{0}^{2}(\rho/\rho_{0})^{2/3}]^{-3/2}
\end{eqnarray}
This expression would have a somewhat simpler form under the assumption that the bulk modulus is pressure independent,that is $B_{0}'=0$. As $B_{0}\neq 0$ the last expression does not tend to infinity for any real material.

More refined results could be obtained by using analytic approximations to the Helmholtz free energy, from which all the thermodynamic potentials can be derived. This is,for example, the approach used in the ANEOS EOS \cite{[6]}. However, ANEOS uses a file of up to 40 parameters for the characterization of materials,which complicates its applications. 
\section{An example}
\label{sec:4}
As a test, the procedure disucssed in this letter was applied to the case of the Barringer crater in Arizona,a well known impact crater. It was assumed that $\bar{V}=7 km/s$,which is the mean value of the measured velocity  of seismic waves in the Earth. Parts of the impactor have been found,and it is known to be an iron-nickel meteorite with $\rho_{1}\cong 8000 kg/m^{3}$. 
Taking that the temperature of the target is $T_{2}=300K$,inserting the known values of the diameter and depth of the crater ($a = b = 1.2 km$, $c = 0.18 km$) and assuming that the radius of the impactor is $50 m$,it follows that,for vertical impact, the impact velocity  was around $13.7 km/s$. These values are in good agreement with those existing in the literature \cite{[7]}. Repeating the same calculation for the impact angle of $\theta = \pi/4$ leads to the value of the impact velocity of $19.4 km/s$. 
This value of the velocity is very close to the value obtained in \cite{[8]}, but their value of the radius of the impactor is $r=40m$. 
\section{Discussion}
In this paper we have presented a simple procedure which gives the possibility of estimating the value of the product of the density and radius of the impactor and the speed of impact: $\rho_{1} r_{1}^{3} (v \cos \theta)^{2}$ in terms of various material parameters of the target and the impactor. The expression for this product has been derived using basic principles of solid state physics, without any special assumption(s) about the materials. In (at least some of the) terrestrial applications the density of the impactor and the inclination of the trajectory can (in principle) be measured or estimated, which gives the possibility of estimating the impact speed if the value of $r_{1}$ can somehow be estimated. 

The shape of the craters has been approximated as a half of a rotating elipsoid,regardless of the composition of the material of the target. As a first approximation, the speed of the seismic waves has been "put in" instead of being calculated from the equation of state of the material of the target. Calculating this speed would have demanded the precise knowledge of the chemical composition of the target materials,and of $B_{0}$ and $B_{0}'$ for them.  
The impact of a projectile into a target leads to heating,and possibly melting and even vaporisation of the target around the impact point. The thermodynamic result of an impact depends on the heat capacity of the material of the target and on the kinetic energy of the  impactor. The heat capacity can be measured, or theoretically estimated, assuming prior knowledge of the chemical composition of the target. Knowledge of the heat capacity is vital for estimates of the temperature to which the target heats in the impact.Therefore, the question is which impacts lead to melting and vaporisation of the material of the target,and which just provoke heating of the target. Even if the material of the target gets partially melted and as such flows away, the dimensions of the crater formed by the impact are determined by the volume of the material pushed away in the impact. This volume is,in turn, determined by the ratio of the kinetic energy of the impactor to the internal energy of some volume of the target. Some details of this problem have been discussed in \cite{[9]}.

The general conclusion of work reported here is that by using simple well known results of solid state physics and a given appoximation of the shape of impact craters, it becomes possible to estimate the value of the product $\rho_{1} r_{1}^{3} (v \cos \theta)^{2}$. This in turn can be used to draw conclusions about the impactors which made various craters on the solid surfaces of objects in the planetary system.  
    
\section{Acknowledgement} 

The preparation of this work was financed by the Ministry of Education,
Science and Technology of Serbia under project 174031. 

\section {Note} 
This paper was submitted to Earth Moon and Planets. It was rejected by one of the editors, without being sent to a referee, on the grounds that the equations are  known and that the link http://impact.ese.ic.ac.uk/ImpactEffects/ offers more data. 
{}
\end{document}